# *i*-FlatCam: A 253 FPS, 91.49 μJ/Frame Ultra-Compact Intelligent Lensless Camera System for Real-Time and Efficient Eye Tracking in VR/AR

Yang Zhao[1], Ziyun Li[2], Haoran You[1], Yonggan Fu[1], Yongan Zhang[1], Chaojian Li[1], Cheng Wan[1], Shang Wu[1], Xu Ouyang[1], Vivek Boominathan[1], Ashok Veeraraghavan[1], and Yingyan Lin[1*]

[1]Rice University, Houston, Texas, USA; [2]Meta, Redmond, Washington, USA; [*]Corresponding author: yingyan.lin@rice.edu

## Abstract

We present a first-of-its-kind ultra-compact intelligent camera system, dubbed *i*-FlatCam, including a lensless camera with a computational (Comp.) chip. It highlights (1) a predict-then-focus eye tracking pipeline for boosted efficiency without compromising the accuracy, (2) a unified compression scheme for single-chip processing and improved frame rate per second (FPS), and (3) dedicated intra-channel reuse design for depth-wise convolutional layers (DW-CONV) to increase utilization. *i*-FlatCam demonstrates the first eye tracking pipeline with a lensless camera and achieves 3.16 degrees of accuracy, 253 FPS, 91.49 μJ/Frame, and 6.7mm×8.9mm×1.2mm camera form factor, paving the way for next-generation Augmented Reality (AR) and Virtual Reality (VR) devices.

## The Proposed *i*-FlatCam System

Eye tracking is an essential human-machine interface modality in AR/VR, requiring stringent efficiency (e.g., >240 FPS and power consumption in milli-watts) and form factor to operate and be fitted in AR/VR glasses [1]. However, existing eye tracking systems are still an order of magnitude slower [2, 3] and require a large form factor due to their lens-based cameras (e.g., 10-20mm in thickness [4]). Hence, this work proposes, develops, and validates an ultra-compact lensless intelligent camera system, *i*-FlatCam (**Fig. 1**), consisting of (1) a lensless camera called FlatCam and (2) a Comp. chip for compact, real-time, and low-power eye tracking for VR/AR.

**The FlatCam** replaces the focal lens of lens-based cameras with a much thinner coded binary mask (<2mm, i.e., 5-10× thinner than lens-based cameras), which encodes the incoming light instead of directly focusing it [4], and its encoded sensing measurements can be decoded [4] to reconstruct scene images.

**The Comp. chip** features a predict-then-focus pipeline that extracts ROIs of only 24% (average) the original images from near-eye cameras [5] for gaze estimation to reduce redundant computations and data movements. Additionally, the temporal correlation across frames is leveraged so that only 5% of the frames require ROIs adjustment over time. These reduce FLOPs of the eye tracking pipeline significantly be 69.49%. To further boosted efficiency, we adopt a unified compression scheme with heterogeneous dataflows for CONV/DW-CONV.

***Chip Architecture***. The Comp. chip (**Fig. 2**) consists of compression-aware modules, 64 PE lines, and memories for the weights and input/output feature maps (IFM/OFM). First, to enable single-chip processing, the weights of both CONV and point-wise (PW)-CONV are compressed via a compression scheme that unifies decomposition, pruning, and quantization, pruning 50% of weights for the gaze estimation model. Second, each PE_line performs 1D row-stationary operations and the 64 PE_lines adopt heterogeneous dataflows (**Fig. 3**) for CONV and DW-CONV to leverage inter- and intra-channel data reuses, respectively, boosting the PE utilization for DW-CONV by 75-87.5%. Third, two levels of memories are adopted for the weights and IFM/OFM.

***Unified Compression for Reduced Storage and Structure Sparsity.*** **Fig. 4** shows the compression algorithm and its hardware supporting modules, enabling 45.7% fewer global buffer (GB) weight accesses and structurally skipped processing. First, weights are stacked as a tall-thin matrix and then decomposed into a small basis matrix (BM) and a large coefficient matrix (CM), with power-of-2 quantization and structure sparsification being enforced in CM. Hence, only a small BM and the non-zero rows of CM (in weight GB) with their run-length encoding indexes (in weight index SRAM) need to be stored, reducing gaze estimation storage by 22×. Second, the restore engine (RE) restores the weights from the BM and CM by using locally stored BM and a shift-and-add unit. Third, the structure sparsity in CM allows row-wise sparsity in CONV and channel-wise sparsity in PW-CONV to skip both corresponding computations and GB weight accesses, leveraging the 2× higher bandwidth for the IFM GB offered by our sequential-write-parallel-read (SWPR) IFM buffer design (Fig. 3, top-right) inserted between the IFM GB and PE_lines.

## Measurement Results

In *i*-FlatCam, the FlatCam's coded binary mask is fabricated in house while the Comp. chip is in 28nm HPC CMOS. **Fig. 5** illustrates the (1) Comp. chip's die photo, (2) performance summary, (3) fabricated mask, (4) FlatCam prototype, and (5) *i*-FlatCam's full system setup. FlatCam, i.e., *i*-FlatCam's camera, has a size of 6.7mm×8.9mm×1.2mm, where the mask is 1.2mm away from the sensor (an advantageous form factor).

**Fig. 7** lists the measured eye tracking results on the industry-standard dataset OpenEDS [5] (see the models' structures in **Fig. 6)**. In accuracy (Fig. 7, top-left), *i*-FlatCam achieves an average angular error of 3.16 degrees, matching the state-of-the-art (SOTA) winners in [5]; In efficiency, compared with the SOTA NN-based eye tracking work [2] and geometric algorithm-based work [3], *i*-FlatCam achieves the required real-time FPS (i.e., >240 FPS), one order of magnitude higher than [2, 3], together with its one order of smaller energy/frame. *i*-FlatCam's energy consumption, including both the FlatCam's sensor and Comp. chip, is 1.59 nJ/pixel, achieving a 2.73× energy saving over [3]; Compared with SOTA vision processors [7, 8], *i*-FlatCam delivers a higher energy efficiency of 0.29-18.9 TOPS/W, with both promising form factor and FPS for eye tracking in AR/VR.

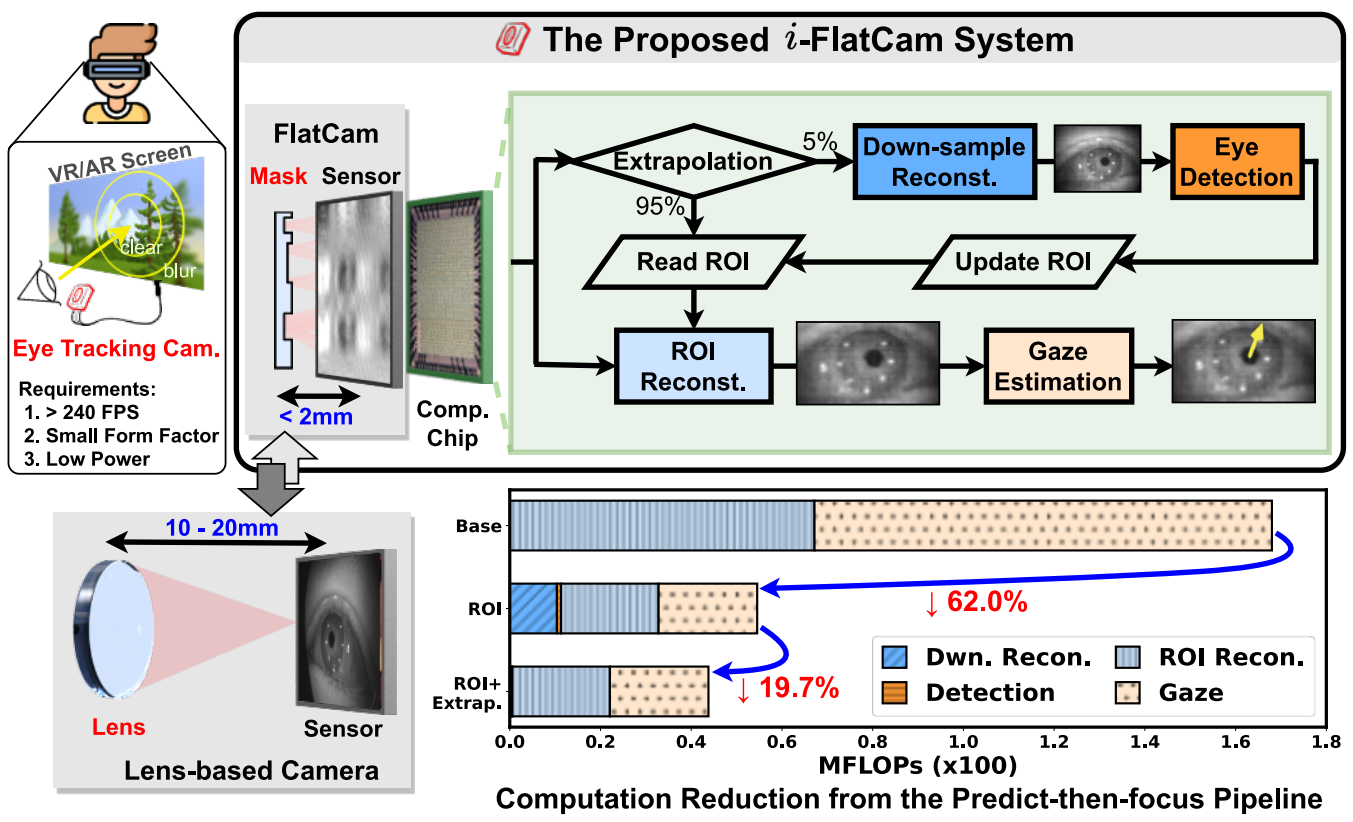

Fig. 1 *i*-FlatCam system with a predict-then-focus pipeline.

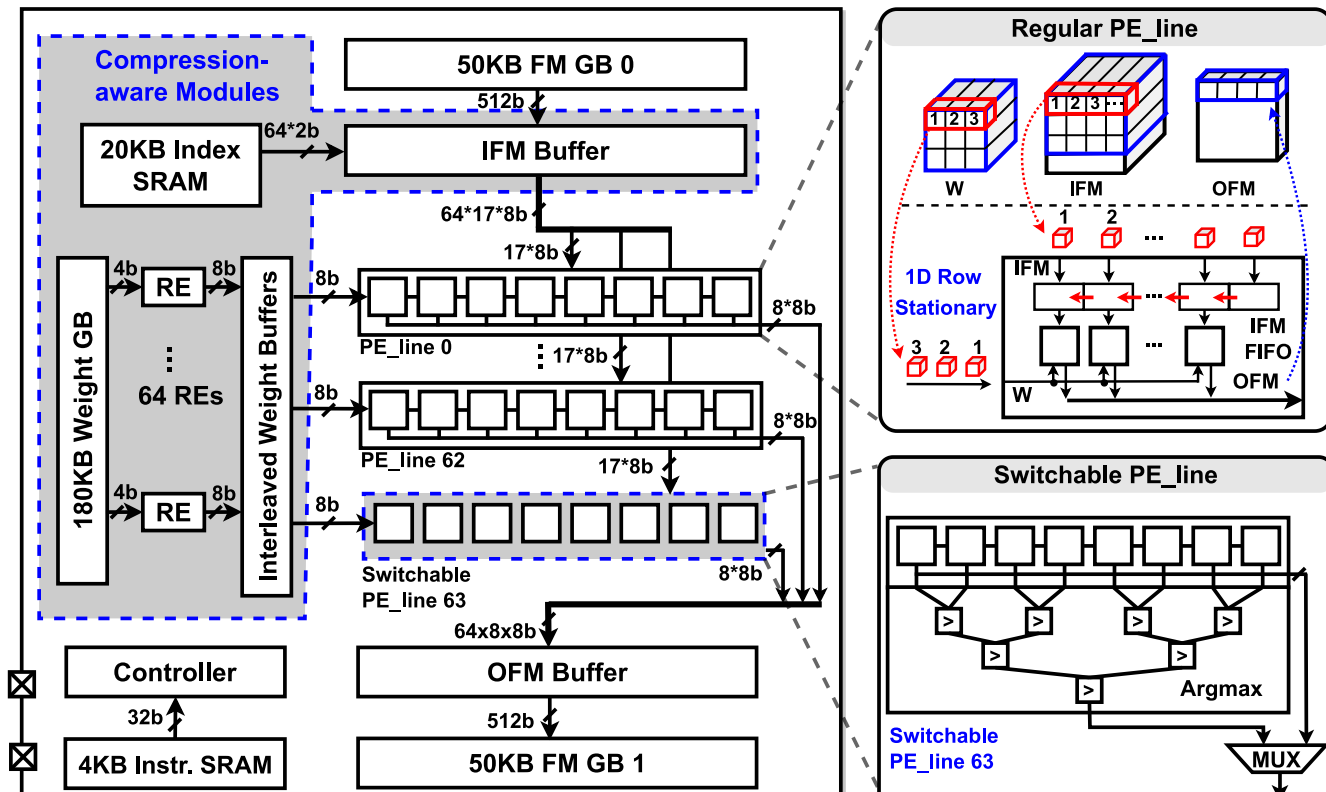

Fig. 2 The block diagram of *i*-FlatCam's Comp. chip.

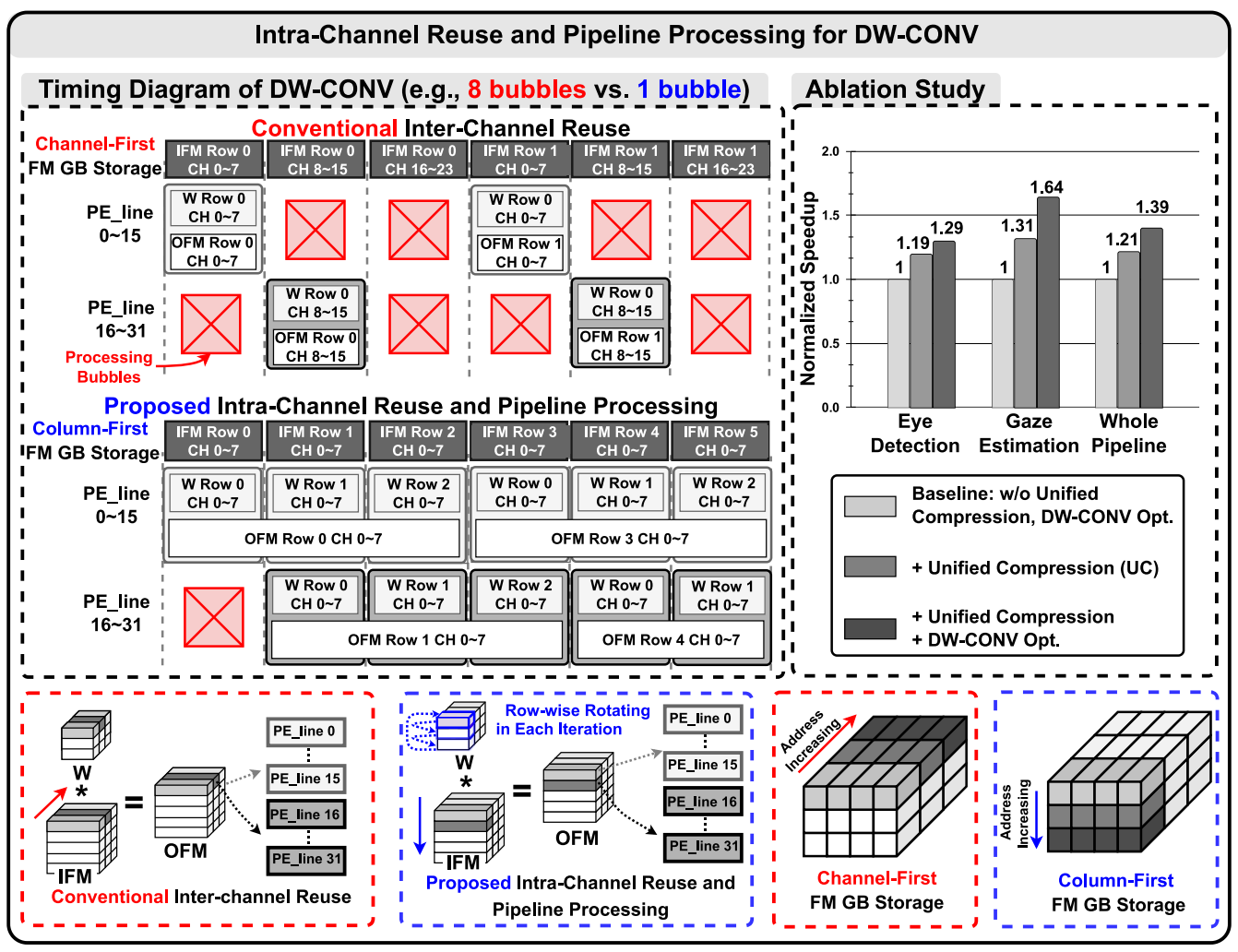

Fig. 3 The proposed intra-channel reuse with pipeline and reconfigurable feature map global buffer storage for DW-CONV.

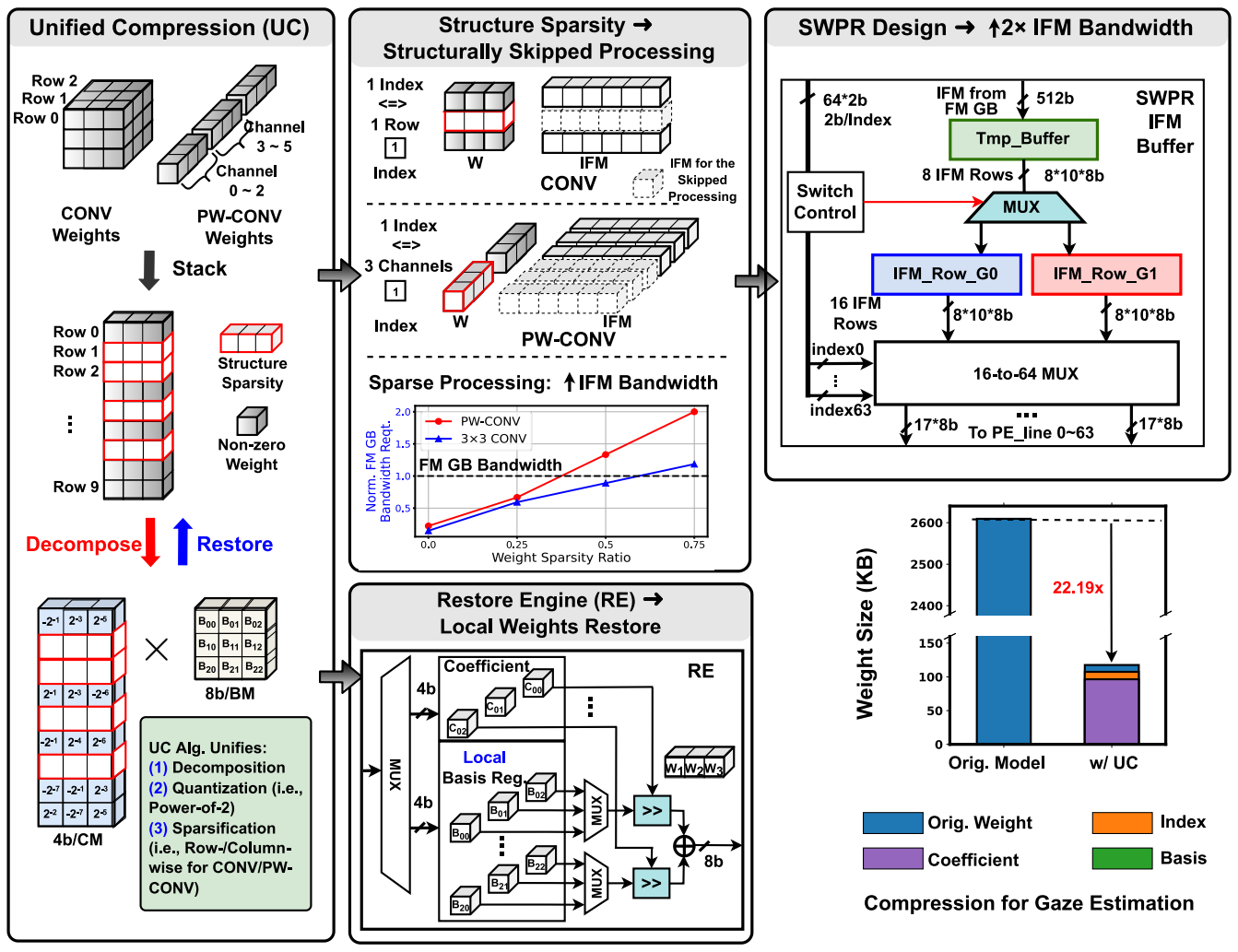

Fig. 4 *i*-FlatCam's unified compression and hardware support.

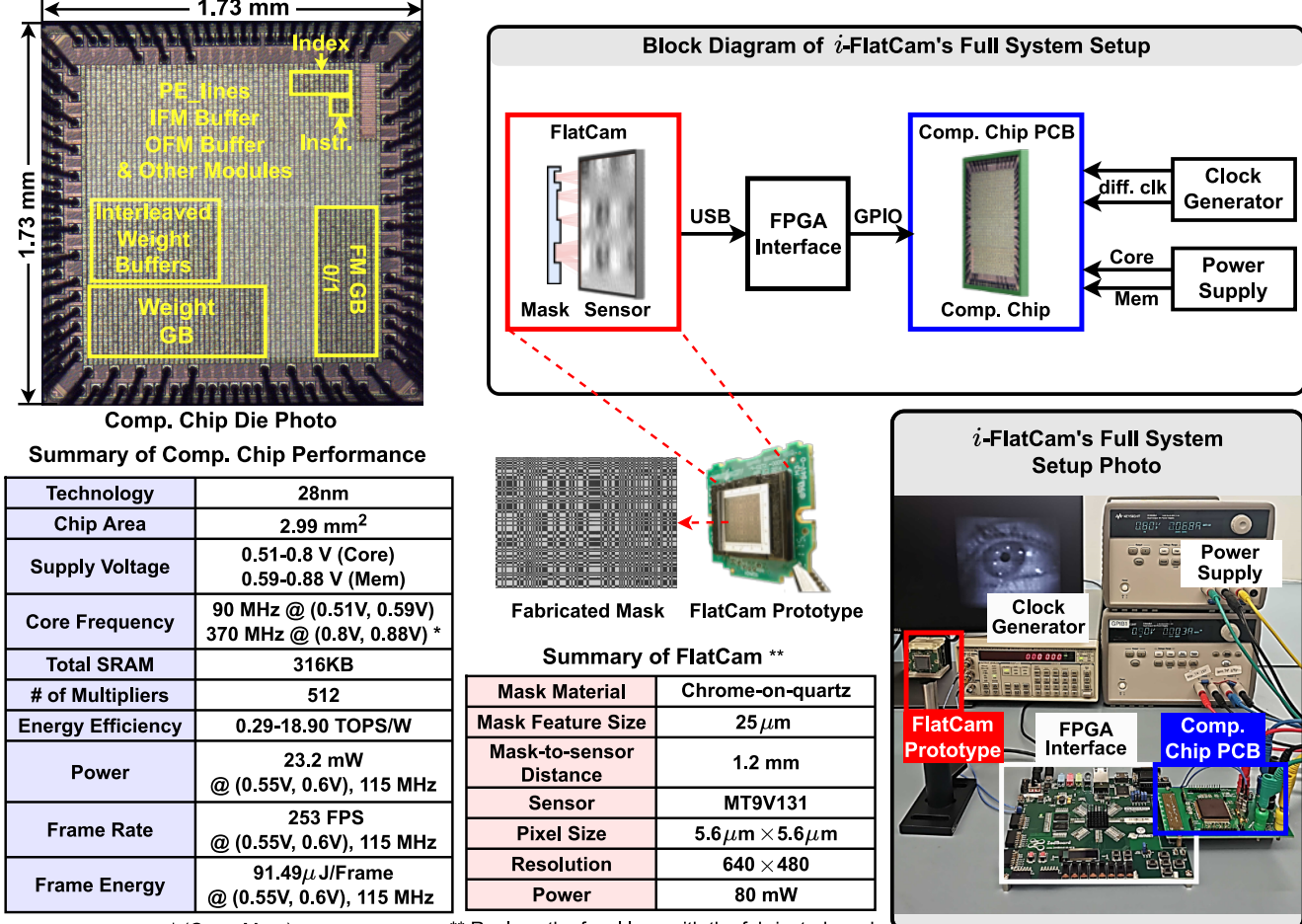

Summary of Comp. Chip Performance

| | |
|---|---|
| Technology | 28nm |
| Chip Area | 2.99 mm$^2$ |
| Supply Voltage | 0.51-0.8 V (Core)<br>0.59-0.88 V (Mem) |
| Core Frequency | 90 MHz @ (0.51V, 0.59V)<br>370 MHz @ (0.8V, 0.88V) * |
| Total SRAM | 316KB |
| # of Multipliers | 512 |
| Energy Efficiency | 0.29-18.90 TOPS/W |
| Power | 23.2 mW<br>@ (0.55V, 0.6V), 115 MHz |
| Frame Rate | 253 FPS<br>@ (0.55V, 0.6V), 115 MHz |
| Frame Energy | 91.49$\mu$J/Frame<br>@ (0.55V, 0.6V), 115 MHz |

* (Core, Mem)

Summary of FlatCam **

| | |
|---|---|
| Mask Material | Chrome-on-quartz |
| Mask Feature Size | 25$\mu$m |
| Mask-to-sensor Distance | 1.2 mm |
| Sensor | MT9V131 |
| Pixel Size | 5.6$\mu$m $\times$ 5.6$\mu$m |
| Resolution | 640 $\times$ 480 |
| Power | 80 mW |

** Replace the focal lens with the fabricated mask.

Fig. 5 The Comp. chip's die photo, performance summary, the fabricated mask, FlatCam prototype, and *i*-FlatCam's full system.

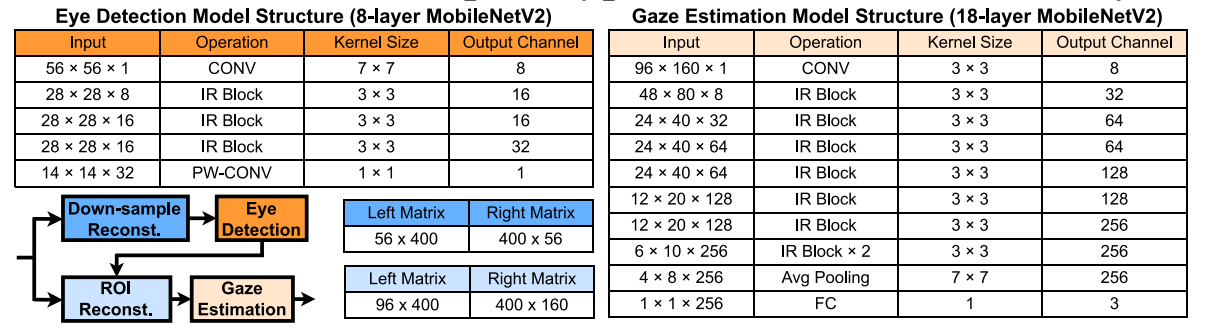

Eye Detection Model Structure (8-layer MobileNetV2)

| Input | Operation | Kernel Size | Output Channel |
|---|---|---|---|
| 56 × 56 × 1 | CONV | 7 × 7 | 8 |
| 28 × 28 × 8 | IR Block | 3 × 3 | 16 |
| 28 × 28 × 16 | IR Block | 3 × 3 | 16 |
| 28 × 28 × 16 | IR Block | 3 × 3 | 32 |
| 14 × 14 × 32 | PW-CONV | 1 × 1 | 1 |

| Left Matrix | Right Matrix |
|---|---|
| 56 x 400 | 400 x 56 |

| Left Matrix | Right Matrix |
|---|---|
| 96 x 400 | 400 x 160 |

Gaze Estimation Model Structure (18-layer MobileNetV2)

| Input | Operation | Kernel Size | Output Channel |
|---|---|---|---|
| 96 × 160 × 1 | CONV | 3 × 3 | 8 |
| 48 × 80 × 8 | IR Block | 3 × 3 | 32 |
| 24 × 40 × 32 | IR Block | 3 × 3 | 64 |
| 24 × 40 × 64 | IR Block | 3 × 3 | 64 |
| 24 × 40 × 64 | IR Block | 3 × 3 | 128 |
| 12 × 20 × 128 | IR Block | 3 × 3 | 128 |
| 12 × 20 × 128 | IR Block | 3 × 3 | 256 |
| 6 × 10 × 256 | IR Block × 2 | 3 × 3 | 256 |
| 4 × 8 × 256 | Avg Pooling | 7 × 7 | 256 |
| 1 × 1 × 256 | FC | 1 | 3 |

Fig. 6. The eye detection and gaze estimation model structure.

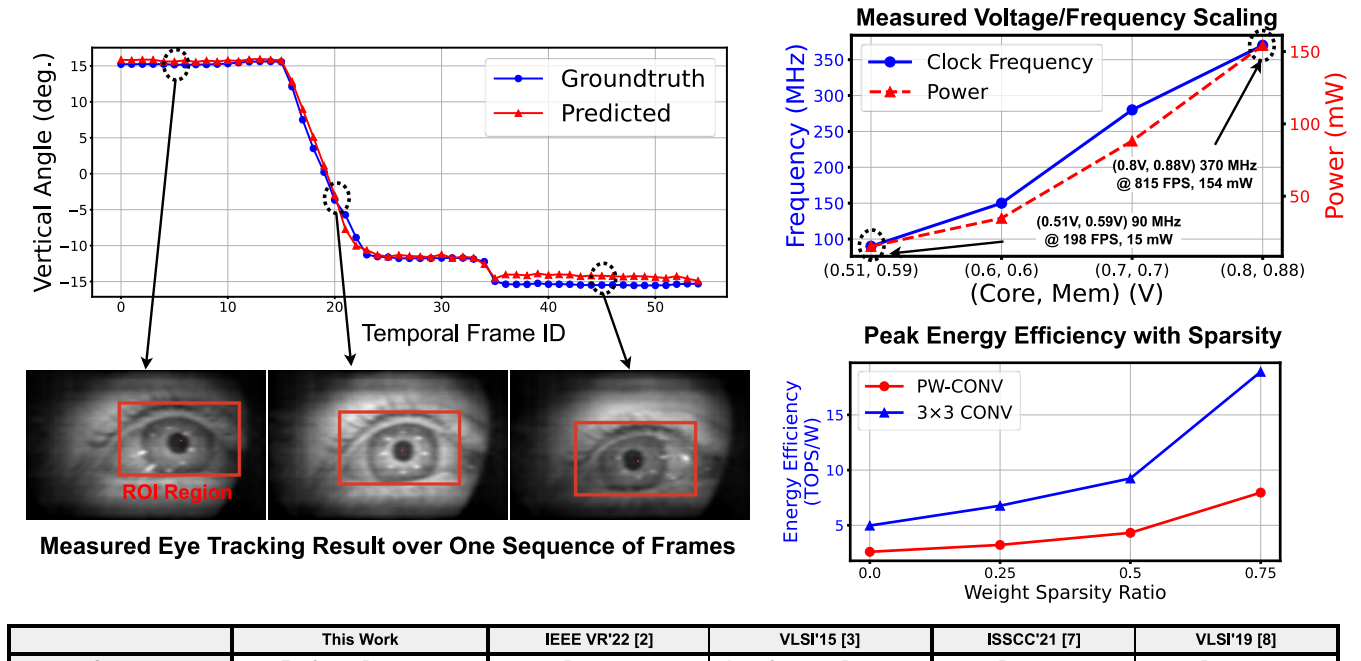

| | | This Work | IEEE VR'22 [2] | VLSI'15 [3] | ISSCC'21 [7] | VLSI'19 [8] |
|---|---|---|---|---|---|---|
| System | | FlatCam + Processor | Processor | Lens Camera + Processor | Processor | Processor |
| Tasks | | Eye Detection<br>Gaze Estimation | Eye Segmentation<br>Gaze Estimation | Pupil Detection<br>Gaze Estimation | Image Classification | Image Classification |
| Algorithms | | DNN (Eye Detection)<br>DNN (Gaze Estimation) | DNN (Eye Segmentation)<br>Regression (Gaze Estimation) | Geometric (Pupil Detection)<br>Regression (Gaze Estimation) | DNN (Image Classification) | DNN (Image Classification) |
| Process (nm) | | 28 | Jetson Xavier | 65 | 28 | 16 |
| Core Area (mm^2) | | 2.99 | | 11.29 | 1.9 | 2.4 |
| Supply Voltage (V) | | 0.51-0.80 (Core)<br>0.59-0.88 (Mem) | | 2.5 (Sensor)<br>1.2 (Digital) | 0.6-0.9 | 0.55-0.80 |
| Frequency (MHz) | | 90-370 | | 50 | 100-470 | 33-480 |
| Memory (KB) | | 316 | | 127 | 206 | 280.6 |
| Bit Precision | | 4/8 (W), 8 (FM) | | 32 (Floating-point) | 8 | 16 |
| Support NN | | Yes | | No | Yes | Yes |
| # of Multipliers | | 512 | | N/A | 864 (Accumulator) | 252 |
| Energy Efficiency (TOPS/W) [2] | | 0.29-18.90 [3] | N/A | 0.03 | 12.62 | 5.06 |
| Resolution | | 640 × 400 | 640 × 400 | 320 × 240 | 224 × 224 | 224 × 224 |
| Frame Rate (FPS) | Each Step | 959-1025 (Reconstruction)<br>5837 (Eye Detection)<br>398 (Gaze Estimation) [1] | N/A | N/A | 40.4 | 91 |
| | Average | 253 [1] | 30 | 30 | | |
| Processor Power | | 23.2 mW [1] | 10~30 W | < 10 mW | 125.8 mW | 16.3-364 mW |
| Processor Frame Energy (μJ/Frame) | | 91.49 [1] | N/A | < 333.33 | 3113.86 | 6020 |
| Mask/Lens | | Customized Mask | Lens | Lens | Lens | Lens |
| Mask/Lens-to-sensor Distance | | 1.2 mm | N/A | ~10-20 mm | N/A | N/A |
| System Energy Consumption (nJ/pixel) [4] | | 1.59 | N/A | 4.34 | N/A | N/A |

1. The performance is tested when running eye tracking application @ 0.55V (Core), 0.60V (Mem), 115MHz.
2. Each MAC computation is considered as 2 operations.
3. The maximum efficiency when running 3x3 kernel weight with 75% row-wise weight sparsity @ 0.51V (Core), 0.59V(Mem), 90MHz.
4. Energy consumption calculation follows [6].

Fig. 7 Measurement results and the comparison with prior works.